\documentclass[%
 reprint,
 amsmath,amssymb,
 aps,
 pra,
]{revtex4-2}

\usepackage{graphicx}
\usepackage{dcolumn}
\usepackage{bm}
\usepackage{physics}
\usepackage{xcolor}
\usepackage[capitalize]{cleveref}
 \usepackage{appendix}

\begin{document}

\preprint{APS/123-QED}

\title{High-Dimensional Entanglement of Photonic Angular Qudits}

\author{Graciana Puentes$^{1,2}$}%
 \email{gpuentes@df.uba.ar}
\affiliation{1-Departamento de Fsica, Facultad de Ciencias Exactas y Naturales,
Universidad de Buenos Aires, Ciudad Universitaria, 1428 Buenos Aires, Argentina\\
2-CONICET-Universidad de Buenos Aires, Instituto de Fsica de Buenos Aires (IFIBA), Ciudad Universitaria,
Buenos Aires, Argentina.
}%

\author{Giacomo Sorelli$^{3}$}%
 
\affiliation{3-Laboratoire Kastler Brossel, Sorbonne Université ENS-Université PSL, CNRS, Collège de France, 4 Place Jussieu, 
F-75252 Paris, France.
}%


\begin{abstract}
We propose a method for generation of entangled photonic states in high dimensions, the so-called qudits, by exploiting quantum correlations of Orbital Angular Momentum (OAM) entangled photons, produced via Spontaneous Parametric Down Conversion. Diffraction masks containing $N$ angular slits placed in the path of twin photons define a qudit space of dimension $N^2$, spanned by the alternative pathways of OAM-entangled photons. We quantify the high-dimensional entanglement of path-entangled photons by the Concurrence, using an analytic expression valid for pure states. We report numerical results for the Concurrence as a function of the angular aperture size for the case of high-dimensional OAM entanglement 
and for the case of high-dimensional path entanglement, produced by $N \times M$ angular slits. Our results provide additional means for preparation and characterization of entangled quantum states in high-dimensions, a fundamental resource for quantum simulation and quantum information
protocols.
\end{abstract}

\maketitle


\section{Introduction}

In recent years, Spontaneous Parametric Down Conversion (SPDC) has become a
fundamental process for generation of entangled photonic states, allowing for preparation of quantum states entangled in several degrees of freedom, such as position and momentum, time and energy, polarization or angular position and
orbital angular momentum (OAM), thus providing for a key resource in fundamentals of quantum physics and quantum information.
In essence, quantum correlations of SPDC photons in a given domain gives rise to interference phenomena resulting from two-photon coherence [1-8].  These  phenomena are routinely used in fundamental tests of quantum physics [9–11] and are a key ingredient for the implementation of quantum communication and information protocols, including quantum teleportation and quantum cryptography [12–14]. Fourier relations link angular position and OAM of photons, leading to angular interference in
the OAM-mode distribution of photons, as they diffract through angular apertures, resulting in two-photon quantum interference [15–22].\\

In this article, we quantify entanglement of such high-dimensional angular qudits, in a scheme in which OAM-entangled photons produced by SPDC are transmitted through multiple angular apertures, in the form of $N \times M$ angular slits in the path of signal and idler photons, which results in path entanglement in a space of dimension $D = N \times M$. Using this scheme, we demonstrate high-dimensional entanglement based on angular-position correlations of down-converted photons. Our results suggests that violations of Bells inequalities in even higher dimensions could in principle be achieved. Moreover, in contrast to previous approaches [16,17] relying on the Fourier limit only, our results shine light on the quantum interpretation, providing new insights. Adding to the novelty of our work, we 
consider the case of asymmetric angular slits $N$ and $M$ for signal and idler, which can lead to high-dimensional angular interference phenomena. We note that linear qudits have  previously been proposed [24], here we propose and characterize angular qudits, which due to their shape can enable generation of entanglement in a larger Hilbert space than
linear qudits. In order to
quantify entanglement, we derive an analytic expression for calculation of the Concurrence, valid for pure states. The results reported here extend the notion of angular qudits to an arbitrary number of angular slits $N \times M$, which not only demonstrates two-photon coherence effects in the angular domain but also provide additional means for preparation and characterization of entangled quantum states in a high-dimensional Hilbert space, which is a fundamental resource for quantum communication [23-40] and quantum information
protocols [41-60].

The article is organized as follows: In section II we introduce the concept of high-dimensional path-entanglement of angular qudits [59]. Next, in section III we present an overview of  angular diffraction in the position basis [60]. In section IV we derive an analytic expression for the Concurrence of high-dimensional qudits, valid for pure states. In  section V we present numerical results in two specific scenarios, (IVA) High dimensional OAM entanglement and (IVB) high-dimensional path entanglement. Finally, in section V we outline the conclusions.

\section{High-Dimensional Angular Qudits}
\label{Sec:concurrence}

 \begin{figure}[h!]

\includegraphics[width=0.5\textwidth]{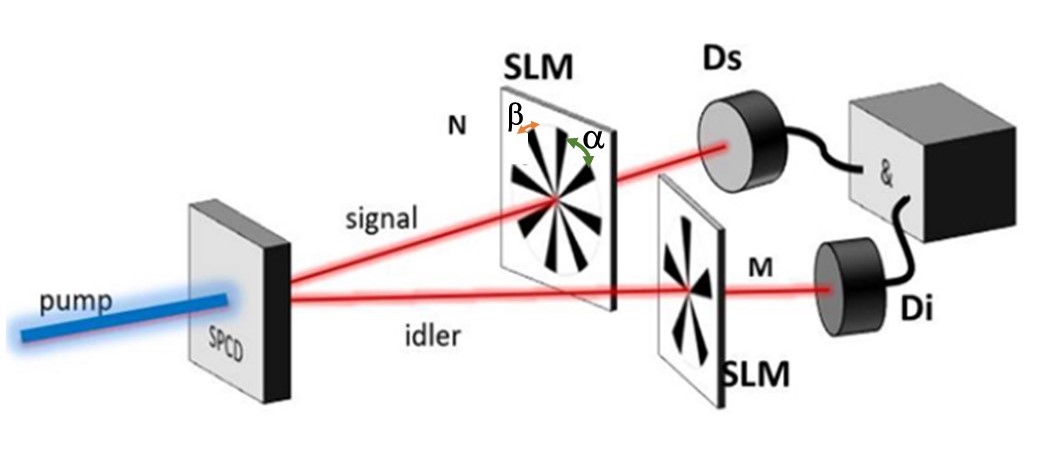}  
\caption{Proposed experimental scheme for generation of high-dimensional OAM-entangled photon pairs via SPDC. Details are in the text.}
\label{Fig:OAM_one_aperture}
\end{figure}

Consider the experimental scheme depicted in Fig. 1. In essence, a Gaussian pump beam produces signal ($s$) and idler ($i$) entangled twin photons by the non-linear process of SPDC. In the simplest scenario of a Gaussian pump  beam with zero OAM ($l=0$), phase matching conditions determine the two-photon down-converted state $|\psi_{\mathrm{0}} \rangle$, which can be expressed in the following form \cite{61}:

\begin{equation}
|\psi_{\mathrm{0}} \rangle= \sum_{l=-N}^{N} c_{l}|l\rangle_{s}|-l\rangle_{i},
\end{equation}

here $s$ and $i$ label signal and idler photons,   $|l\rangle$ refers to the OAM eigen-mode of order $l$, and $D=2N+1$ is the dimension of the OAM Hilbert space under consideration. Such OAM modes are characterized by an azymmuthal phase front typically expressed as $e^{il\phi}$. The expansion coefficients $|c_{l}|^2$ determine the probability of generating photon pairs in a given OAM mode of order $l$. For $|\psi_{0} \rangle$ to represent a quantum state, the normalization condition imposes $\sum_{l=-N}^{N}|c_{l}|^2=1$. Subsequently, signal and idler photons are transmitted through $N $ angular slits, as shown in Fig. 1. The transmission functions $A_{j,n}$ of the individual angular
slits are given by \cite{61}: 

\begin{equation}
A_{j,n}(\phi_{j})= 1 \hspace{0.2cm} \mathrm{if}  \hspace{0.2cm} n\beta -\alpha/2 \leq \phi_{j} \leq n \beta + \alpha/2  \hspace{0.2cm} \mathrm{else}  \hspace{0.2cm} 0,
\end{equation}

where $n=0,...,N-1$ is the angular slit label,  $\alpha$ represents the aperture of the angular slits, and $\beta$ represents the separation between consecutive angular slits. For the simplest case $N=2$ slits, we recover the results reported in \cite{56}. Considering $N$ slits in both arms, there are in principle $N^2$  alternative pathways by which the down-converted photons can
pass through the apertures and get detected in coincidence
at single-photon avalanche detectors $D_{s}$ and $D_{i}$. The $N^2$ alternative paths, here labelled by the sub-index $q=1,..., N^2$, can be expressed as the outer product of the sub-spaces corresponding to each photon $(s,i)$ passing through the slits  $n=0,...,N-1$, respectively, in the following form \cite{61}: 

\begin{eqnarray}
& &|s,0\rangle \otimes \{|i,0\rangle, |i,1\rangle,..., |i,N-1\rangle\};\nonumber\\
& &|s,1\rangle \otimes \{|i,0\rangle, |i,1\rangle,..., |i,N-1\rangle\};....\nonumber \\ 
& &|s,N-1\rangle \otimes \{|i,0\rangle, |i,1\rangle,..., |i,N-1\rangle\}. \nonumber\\
\end{eqnarray}

Due to quantum correlations between the position of the two photons, only paths of the form $|i,n\rangle|s,n\rangle$ will have a significant contribution. In this notation, the twin-photon diffracted state by the $k$-th slit, with OAM momentum $l$, can be expressed as \cite{61}:

\begin{equation}
| \psi_{l}^{k} \rangle = \frac{\alpha}{2\pi}\sum_{l'} e^{-i(l'-l)\beta k} \mathrm{sinc}[\frac{\alpha(l'-l)}{2}]|l'\rangle.
\end{equation}

In the most general case, the overlap between the diffracted states by slits $k$ and $j$, with OAM labels $m$ and $l$, results in:

\begin{eqnarray}
\langle\psi_{m}^{j} | \psi_{l}^{k} \rangle & = & (\frac{\alpha}{2\pi})^2 \sum_{l'} \sum_{l''}  e^{-i(l'-l)\beta k} e^{i(l''-m)\beta j} \times \\ \nonumber
&  &\mathrm{sinc}[\frac{\alpha(l'-l)}{2}]  \mathrm{sinc}[\frac{\alpha(l''-m)}{2}] \langle l'' |l'\rangle,\\ \nonumber
\end{eqnarray}

where $\langle l'' |l'\rangle= \delta_{l'',l'}$ due to orthogonality of OAM modes.\\

For the simple case of a single slit in both arms, we have $j=k=1$, the mode overlap $ b_{ml}=\langle \psi_{m} | \psi_{l} \rangle$ can be expressed as:

\begin{eqnarray}
b_{ml} & = & (\frac{\alpha}{2\pi})^2 \sum_{l'}  e^{i(l-m)\beta} \times  \\ \nonumber
&  &\mathrm{sinc}[\frac{\alpha(l'-l)}{2}]  \mathrm{sinc}[\frac{\alpha(l'-m)}{2}].\\ \nonumber
\end{eqnarray}

\section{High-dimensional spatial mode entanglement and diffraction}

In this Section we derive expressions for twin photons angular diffracted states in the position basis.
 As readily anticipated in Section II, we focus on $D-$dimensional ($D = 2N+1$) entangled biphoton pure states of the form \cite{62}
\begin{equation}
    \ket{\psi_0} = \sum_{l=-N}^N c_l \ket{l}\ket{-l},
    \label{psi0}
\end{equation}
where 
\begin{equation}
    \ket{l} = \int d^2 {\bf r} u_l({\bf r}) \ket{{\bf r}}
\end{equation}
are single photon states in the spatial modes defined by the orthonormal transverse functions $u_l({\bf r})$, i.e. $\int d^2{\bf r} u_{l^\prime}^*({\bf r})u_l({\bf r})= \delta_{l,l^\prime}$, and $\sum_{l=-N}^{N} |c_l|^2 =1$ to ensure normalization \cite{62,63}.\\

We are interested in how the entanglement of such states is modified by diffraction of the two photons on two independent opaque screens of different shapes and forms. Moreover, we focus on the case where both photons are detected after the diffracting screens. In this post-selection scenario, entanglement is only modified by the change of the spatial profile of the modes occupied by the photons due to diffraction on the screens. 
Under these assumptions, the quantum state of the diffracted photons can be written \cite{62,63} 
\begin{equation}
    \ket{\psi} = \sum_{l=-N}^{N} \tilde{c}_l \ket{\psi_l}\ket{\psi_{-l}},
\end{equation}
with 
\begin{equation}
    \ket{\psi_l} =  \int d^2 {\bf r} \psi_l({\bf r}) \ket{{\bf r}},
    \label{diffracted_single_photon}
\end{equation}
where the modes $\psi_l({\bf r})$ are the images of the modes $u_l({\bf r})$ that can be computed using standard diffraction theory \cite{62}, and $\tilde{c}_l = c_l/\sqrt{\mathcal{N}}$ with $\mathcal{N}$
a renormalization constant needed to compensate for the non orthogonality of the diffracted modes $\psi_l({\bf r})$, which we can express as \cite{62,63}
\begin{equation}
    \mathcal{N} = \sum_{lk=-N}^N c^*_l c_k b_{lk}b_{-l-k},
\end{equation}
with
\begin{equation}
    b_{lk} = \int d^2 \psi_l^*({\bf r}) \psi_k({\bf r}),
    \label{mutual_overlap}
\end{equation}
the mutual overlaps between pairs of diffracted modes \cite{62}. We note that since the single photon states $\ket{\psi}_l$ are fully determined by the modes they occupy, we also have that $\braket{\psi_l}{\psi_k} = b_{lk}$.

We now set to characterize the entanglement of such states developing a high-dimensional generalization of the method presented in \cite{61,62}. 
Following this path, we quantify the entanglement of the diffracted state using the Concurrence, that for pure states in arbitrary dimensions can be expressed as a function of the purity $\Tr \rho^2$ of the reduced density matrix $\rho = \Tr_2 \ket{\psi}\bra{\psi}$ \cite{62} 
\begin{equation}
    C(\ket{\psi}) = \sqrt{2\left(1 - \Tr \rho^2 \right)}.
    \label{concurrence}
\end{equation}
The reduced density matrix can be computed using the expression~\eqref{diffracted_single_photon} for the quantum state of single photons in the diffracted modes and the mutual overlap~\eqref{mutual_overlap}
\begin{equation}
    \rho = \int d^2 {\bf r}_2 \braket{{\bf r}_2}{\psi}\braket{\psi}{{\bf r}_2} = \sum_{kl= -N}^N \tilde{c}^*_l\tilde{c}_k b_{-l-k}\ket{\psi_l} \bra{\psi_k}.
\end{equation}
In a similar fashion, we can determine the purity of the reduced density matrix
\begin{align}
    \tr \rho^2 &= \int d^2 {\bf r}_1 \bra{{\bf r}_1}\rho^2\ket{{\bf r}_1} \nonumber \\ 
    &= \sum_{l,k,p,q = -N}^N \tilde{c}_l^*\tilde{c}_k \tilde{c}_p^*\tilde{c}_q b_{-l-k} b_{-p-q} b_{kp} b_{lq}, \label{purity} \\
    &= \frac{\sum_{l,k,p,q = -N}^N c_l^*c_k c_p^*c_q b_{-l-k} b_{-p-q} b_{kp} b_{lq}}{(\sum_{lk=-N}^N c^*_l c_k b_{lk}b_{-l-k})^2} \nonumber
\end{align}
which fully determines the Concurrence Eq. ~\eqref{concurrence}. 
From Eq.~\eqref{purity}, we see that the entanglement of diffracted biphoton states if fully characterize by the entanglement of the input state (encoded in the coefficients $c_{l}$), and by the diffraction induced overlaps of the states $\ket{\psi_l}$ (quantified by the coefficients $b_{lk}$).

\section{Examples}
In the following, we will apply the theoretical framework presented in Sections ~\ref{Sec:concurrence} and III to two particular examples. In particular, in Sec.~\eqref{Sec:HD_OAM}, we will consider high-dimensional OAM-entangled biphotons diffracted on screens containing a single angular aperture, while in Sec.~\eqref{Sec:path_entanglement} we will consider initial biphotons entangled in few OAM modes impinging on screens containing multiple apertures.

\begin{figure}[h!]
\includegraphics[width=0.5\textwidth]{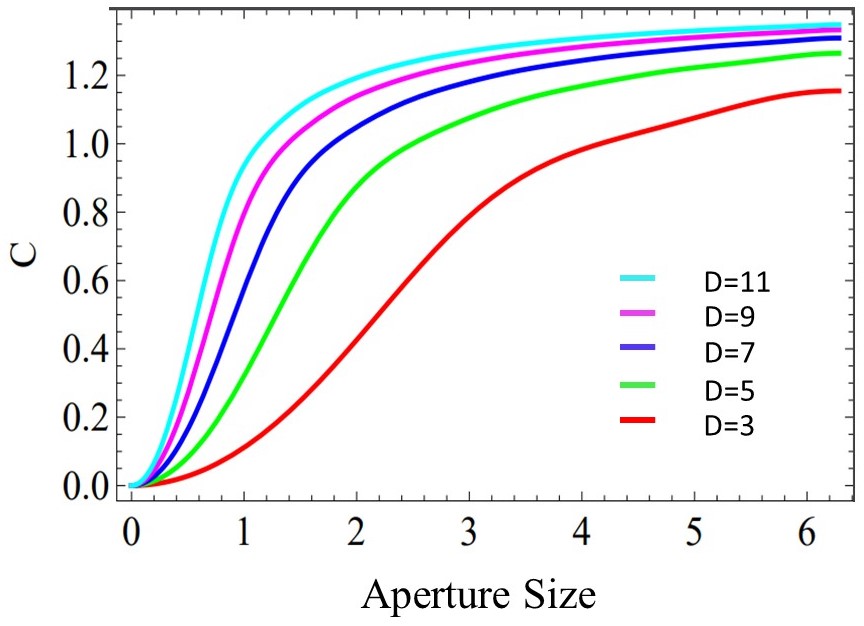}
\caption{Concurrence vs. angular aperture size for different OAM subpsace dimension ($D= 2N+1$) as indicated in the label. Details are in the text.}
\label{Fig:OAM_one_aperture}
\end{figure} 

\subsection{High-dimensional OAM entanglement on angular apertures}
\label{Sec:HD_OAM}
We now specialize to the study of maximally entangled states, i.e we set $c_l = 1/\sqrt{2N+1}\; \forall l$ in Eq.~\eqref{psi0}. Moreover, we assume entanglement in spatial modes carrying OAM, which for simplicity we assume to be fully characterized by their helical phase front so that we have
\begin{equation}
    \ket{l} = \frac{1}{\sqrt{2\pi}} \int_{-\pi}^{\pi} d \phi e^{i l \phi} \ket{\phi}, 
\end{equation}
which after transmission through an angular aperture of size $\alpha$ becomes 
\begin{equation}
    \ket{\psi_l} = \frac{1}{\sqrt{\alpha}}\int_{-\alpha/2}^{\alpha} d \phi e^{i l \phi} \ket{\phi}.
    \label{diff_OAM}
\end{equation}
Using Eq.~\eqref{diff_OAM}, we can calculate the mutual overlaps \eqref{mutual_overlap} which results into 
\begin{equation}
    b_{lk} = \frac{1}{\alpha}\int_{-\alpha/2}^{\alpha} d \phi e^{i(k-l)\phi} = {\rm sinc}\left[\frac{(l-k)\alpha}{2}\right],
    \label{b_OAM}
\end{equation}
with ${\rm sinc}[x] = \sin[x]/x$. 
Eq.~\eqref{b_OAM} implies that $b_{lk} = b_{-l-k}$, accordingly in this case the purity \eqref{purity} can be rewritten as
\begin{equation}
    \Tr \rho^2 = \frac{\sum_{lkpq=-N}^N b_{lk}b_{pq}b_{lp}b_{kq}}{\left(\sum_{lk=-N}^N b_{lk}^2\right)^2}.
    \label{OAM_purity}
\end{equation}
Combining Eqs.~\eqref{OAM_purity} and \eqref{b_OAM} into Eq.~\eqref{concurrence} we can compute the concurrence of the diffracted state, which is plotted in Fig.~\ref{Fig:OAM_one_aperture}.

\subsection{High-dimensional path-entanglement on angular apertures}
\label{Sec:path_entanglement}

\begin{figure}[b!]
\includegraphics[width=0.5\textwidth]{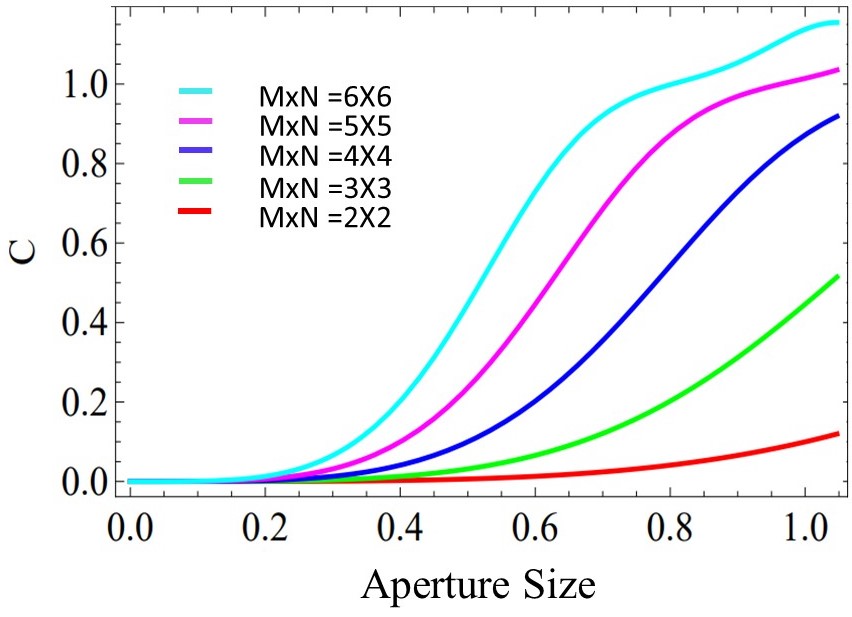}
\caption{Concurrence vs. angular aperture size for different path dimensions  ($M \times N$) and aperture sizes in the range $[0, 2 \pi/N_{\mathrm{max}}]$ with $N_{\mathrm{max}}=6$. Details are in the text.}
\end{figure} 

We now consider an alternative scenario for generation of high-dimensional entanglement relying on path-entangled modes generated by multiple angular apertures in the path of signal and idler photons  ($s,i$). For simplicity, we consider twin photons initially prepared in a OAM qubit state of the form: 
\begin{equation}
|\psi_{0} \rangle=|l_0\rangle_{s} |-l_0 \rangle_{i}. 
\end{equation}
When masks with
($N,M$) angular slits for ($s,i$) photons are placed in each spatial mode (seen Fig. 1), the biphoton state immediately after the apertures can be expressed as \cite{Machado:2019}:
\begin{equation}
|\psi' \rangle \propto \sum_{k=-\frac{(N-1)}{2}}^{\frac{
(N-1)}{2}} \sum_{k'=-\frac{(M-1)}{2}}^{\frac{
(M-1)}{2}} \phi(x_{s},x_{i},z) |\psi^{k} \rangle_{s} |\psi^{k'}\rangle_{i}, 
\end{equation}

where $x_{s,i}$ denote the position of  signal and idler photons at the crystal plane ($z$). $\phi(x_{s},x_{i},z)$ indicates the biphoton amplitude at $z$, which can be regarded as a product of the pump transverse amplitude and phase matching function \cite{Machado:2019}. For simplicity, in what follows we consider the biphoton amplitude to be approximately constant. $|\psi^{k,k'} \rangle_{s,i}$ describe the quantum state of signal and idler photons diffracted by angular apertures $(k,k')$, respectively.\\
\begin{figure}[t!]
\includegraphics[width=0.5\textwidth]{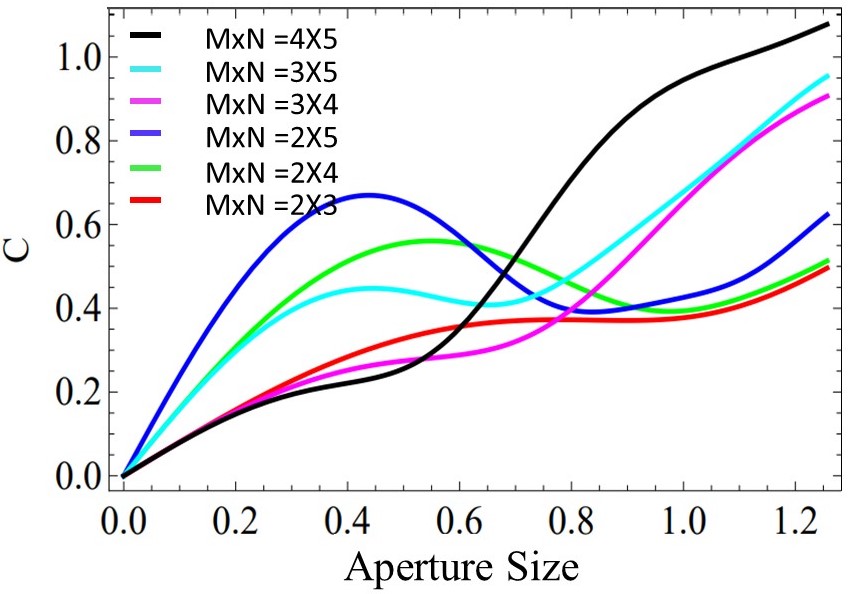}
\caption{Concurrence vs. angular aperture size for different path dimensions  ($M \times N$) and aperture sizes in the range $[0, 2 \pi/N_{\mathrm{max}}]$ with $N_{\mathrm{max}}=5$. Details are in the text.}
\end{figure} 

In the notation introduced in Section II, for an initial state of the form $|l_0\rangle_{s}|-l_0\rangle_{i}$, the quantum state of ($s,i$) photons diffracted by angular apertures ($k,k'$) respectively, can be expressed as:

\begin{equation}
|\psi_{l_0}^{k} \rangle_{s} = \frac{\alpha}{2 \pi} \sum_{l'} e^{-i(l'-l_0)\beta k} \mathrm{sinc}[\frac{\alpha(l'-l_0)}{2}]|l'\rangle_{s},
\end{equation}

\begin{equation}
|\psi_{-l_0}^{k'} \rangle_{i} = \frac{\alpha}{2 \pi} \sum_{l''} e^{-i(l''+l_0)\beta k'} \mathrm{sinc}[\frac{\alpha(l''+l_0)}{2}]|l''\rangle_{i}.
\end{equation}

The biphoton path-entangled state diffracted by $N \times M$ angular slits results in:

\begin{eqnarray}
|\psi' \rangle & = &  \sum_{k=-\frac{(N-1)}{2}}^{\frac{
(N-1)}{2}} \sum_{k'=-\frac{(M-1)}{2}}^{\frac{
(M-1)}{2}} (\frac{\alpha}{2\pi})^2 \times \\ \nonumber
   &  & \sum_{l'} \sum_{l''}  e^{-i(l'-l_0)\beta k} e^{-i(l''+l_0)\beta k'} \times \\ \nonumber
&  &\mathrm{sinc}[\frac{\alpha(l'-l_0)}{2}]\mathrm{sinc}[\frac{\alpha(l''+l_0)}{2}] |l'\rangle_{s}|l''\rangle_{i}.\\ \nonumber
\end{eqnarray}

\begin{figure}[h!]
\includegraphics[width=0.5\textwidth]{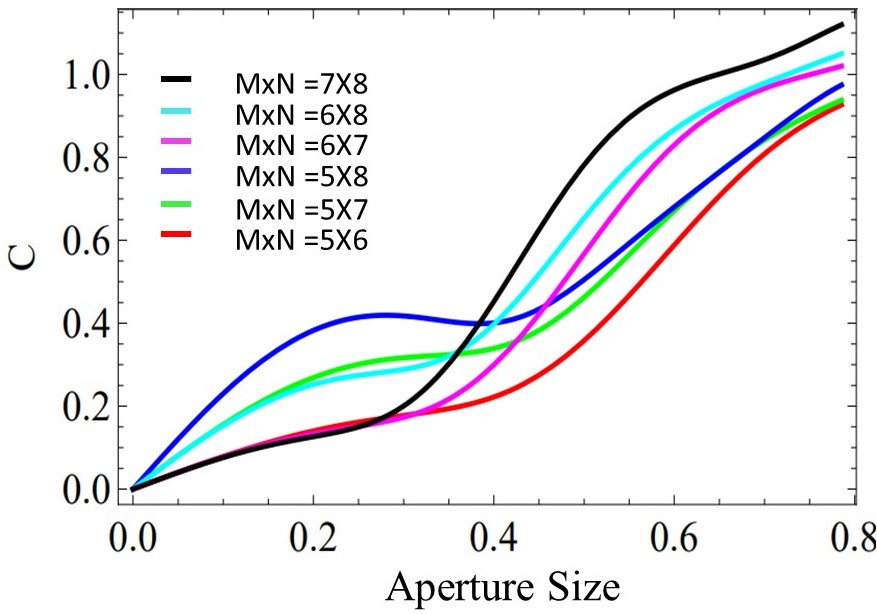}
\caption{Concurrence vs. angular aperture size for different path dimensions  ($M \times N$) and aperture sizes in the range $[0, 2 \pi/N_{\mathrm{max}}]$ with $N_{\mathrm{max}}=8$. Details are in the text.}
\end{figure} 

A compact expression for the biphoton diffracted state can be casted in the form:

\begin{equation}
|\psi'\rangle = \sum_{l'}\sum_{l''} c_{l',l''}|l' \rangle_{s}|l'' \rangle_{i},
\end{equation}

where the coefficients in the summation $c_{l',l''}$ are given by:

\begin{eqnarray}
 c_{l',l''} &=& \sum_{k=\frac{-(N-1)}{2}}^{\frac{
(N-1)}{2}} \sum_{k'=\frac{-(M-1)}{2}}^{\frac{
(M-1)}{2}} (\frac{\alpha}{2\pi})^2 \times \\ \nonumber
   &  &   e^{-i(l'-l_0)\beta k} e^{-i(l''+l_0)\beta k'} \times \\ \nonumber
&  &\mathrm{sinc}[\frac{\alpha(l'-l_0)}{2}]\mathrm{sinc}[\frac{\alpha(l''+l_0)}{2}].   
\end{eqnarray}

The Concurrence for the path-entangled biphoton states can be derived considering the generalized overlap ($b_{lm}$) of the form:
\begin{equation}
 b_{lm}=\sum_{k,k'} \int_{-\pi}^{\pi} d\phi e^{i(m-l)\phi} A_{k}A_{k'},   
\end{equation}
where $A_{k,k'}$ describe the angular aperture functions introduced in Section II.
 
Figures 3, 4, and 5 present numerical simulations of the Concurrence ($C$) vs Aperture Size (in radians), for different angular slit dimensions $M \times N$ for signal and idler photons, respectively. For the case of $N$  and $M$ angular slits the maximum angular aperture size per slit is $2 \pi /N$ and $2 \pi /M$, respectively. Taking this limit into consideration, we performed numerical simulations for aperture sizes in the range $2 \pi /N_{\mathrm{max}}$, where $N_{\mathrm{max}}$ is the maximum number of slits between the values of $N$ and $M$ considered in the specific numerical simulations. All angular dimensions considered are experimentally feasible in view of the resolution of state-of-the-art Spatial Light Modulators (SLMs) \cite{61}. Different curves in Fig. 3 correspond to symmetric path-dimensions $M \times N$ given by $2 \times 2$, $3 \times 3$, $4 \times 4$, $5 \times 5$, and $6 \times 6$ angular silts, for aperture sizes in the range $[0,2 \pi/N_{\mathrm{max}}]$, with $N_{\mathrm{max}}=6$. Different curves in Fig. 4 correspond to asymmetric path-dimensions $M \times N$ given by $2 \times 2$, $2 \times 4$, $2 \times 5$, $3 \times 4$, $3 \times 5$ and $4 \times 5$ angular silts, for aperture sizes in the range $[0,2 \pi/N_{\mathrm{max}}]$, with $N_{\mathrm{max}}=5$. Different curves in Fig. 5 correspond to asymmetric path-dimensions $M \times N$ given by $5 \times 6$, $5 \times 7$, $5 \times 8$, $6 \times 7$, $6 \times 8$ and $7 \times 8$ angular silts, for aperture sizes in the range $[0,2 \pi/N_{\mathrm{max}}]$, with $N_{\mathrm{max}}=8$. Interestingly, within our approximation, for a sufficiently large number of angular apertures ($M \times N$) it is possible to reach the same amount of entanglement as with high-dimensional OAM. A study of the Concurrence in the entire aperture size range $[0,2\pi]$ is presented in Appendix A.

\section{Discussion}

We presented a method to generate entangled photonic states in high-dimensional quantum systems, the so-called qudits, by exploiting quantum correlations of OAM-entangled photons produced by the non-linear process of Spontaneous Parametric Down Conversion. Diffraction masks containing $N$ angular slits in the path of twin photons define a qudit space
of dimension $N^2$, spanned by the alternative pathways of the entangled photons. We quantify the entanglement of  path-entangled photons by an explicit calculation of the Concurrence, valid for pure states. We reported numerical results for the Concurrence as a function of the angular aperture size for the case of high-dimensional OAM-entangled
photons and for the case the case of high-dimensional entanglement produced by $N \times M$ angular slits. Interestingly, within our approximation, it is possible to reach the same amount of entanglement using either high-dimensional OAM-entangled photons or path-entangled photons. Our results shine light into the fundamental quantum aspects of two-photon angular interference, and provide alternative means for preparation entangled quantum states in high-dimensions, a fundamental resource for quantum information and quantum simulation protocols [25-60].\\

\section{Acknowledgements}
The authors gratefully acknowledge Sonja Franke-Arnold for helpful discussions. G.P. acknowledges financial support via PICT Startup. \\

\section*{Appendix A}

In order to obtain a quantitative comparison between the amount of entanglement that can be attained using path-entangled photons vs. OAM-entangled photons, we look for an asymptotic trend in the Concurrence. To find such asymptotic trend for the Concurrence vs. the aperture size  in this context, we performed simulations in the entire range $[0, 2 \pi]$. Numerical results are displayed in Figs. A1, A2 and A3. These results can be understood by considering that angular apertures are periodic functions of the aperture size. More interestingly, within our approximation, it is possible to reach the same amount of entanglement by using either OAM-entangled  or path-entangled photons as a resource.

\renewcommand{\thefigure}{A1}
\begin{figure}
\includegraphics[width=0.5\textwidth]{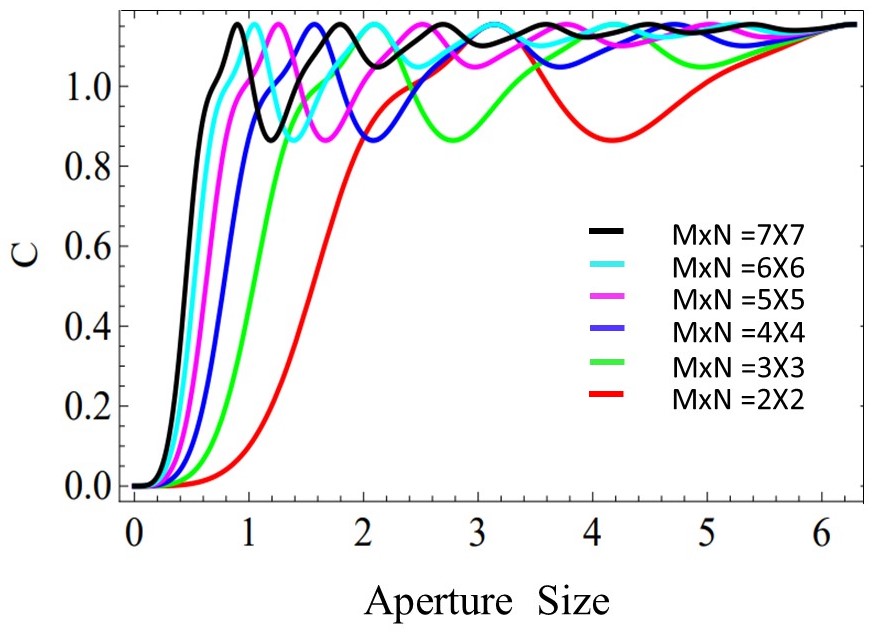}
\caption{Concurrence vs. angular aperture size for different path dimensions  ($M \times N$) and aperture sizes in the range $[0, 2 \pi]$  (see text for details).} 
\end{figure}

\renewcommand{\thefigure}{A2}
\begin{figure}[t!]
\includegraphics[width=0.5\textwidth]{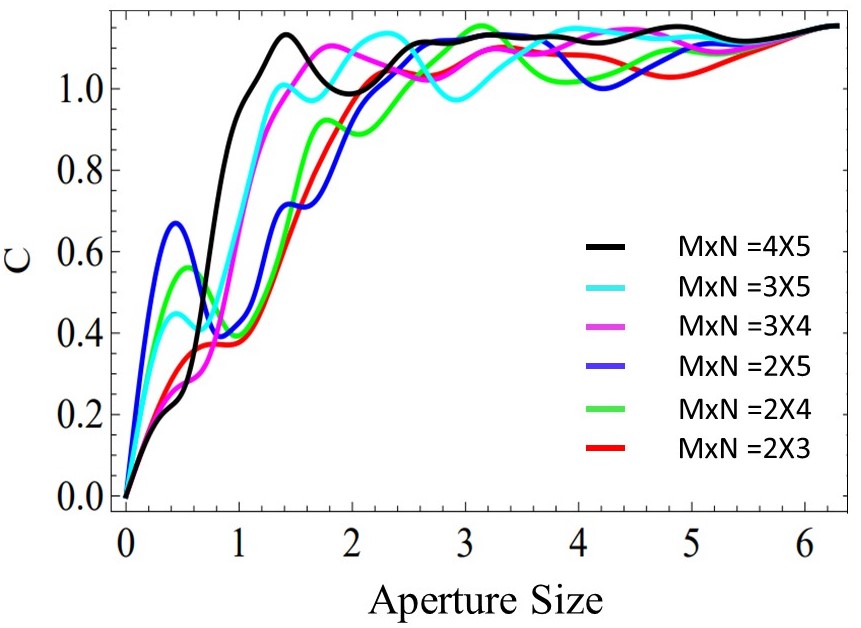}
\caption{Concurrence vs. angular aperture size for different path dimensions  ($M \times N$) and aperture sizes in the range $[0, 2 \pi]$  (see text for details).}
\end{figure}

\renewcommand{\thefigure}{A3}
\begin{figure}[h!]
\includegraphics[width=0.5\textwidth]{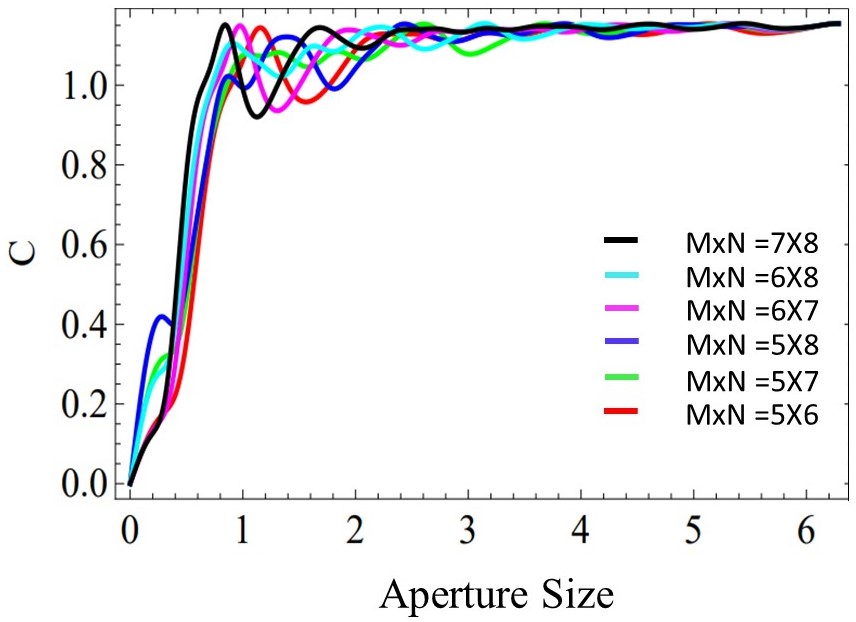}
\caption{Concurrence vs. angular aperture size for different path dimensions  ($M \times N$) and aperture sizes in the range $[0, 2 \pi]$  (see text for details).}
\end{figure}

\end{document}